\documentstyle[prd,aps,preprint,tighten,epsfig]{revtex}

\begin{document}

\draft

\title{Nearly Tri-bimaximal Neutrino Mixing and CP Violation \\
from $\mu$-$\tau$ Symmetry Breaking}
\author{{\bf Zhi-zhong Xing} \thanks{E-mail: xingzz@mail.ihep.ac.cn},
~ {\bf He Zhang} \thanks{E-mail: zhanghe@mail.ihep.ac.cn}, ~ {\bf
Shun Zhou} \thanks{E-mail: zhoush@mail.ihep.ac.cn}}
\address{
CCAST (World Laboratory), P.O. Box 8730, Beijing 100080,
China \\
and
Institute of High Energy Physics, Chinese Academy of Sciences, \\
P.O. Box 918, Beijing 100049, China}

\maketitle

\begin{abstract}
Assuming the Majorana nature of massive neutrinos, we generalize
the Friedberg-Lee neutrino mass model to include CP violation in
the neutrino mass matrix $M^{}_\nu$. We show that a favorable
neutrino mixing pattern (with $\theta^{}_{12} \approx 35.3^\circ$,
$\theta^{}_{23} = 45^\circ$, $\theta^{}_{13} \neq 0^\circ$ and
$\delta = 90^\circ$) can naturally be derived from $M^{}_\nu$, if
it has an approximate or softly-broken $\mu$-$\tau$ symmetry. We
point out a different way to obtain the nearly tri-bimaximal
neutrino mixing with $\delta = 0^\circ$ and non-vanishing Majorana
phases. The most general case, in which all the free parameters of
$M^{}_\nu$ are complex and the resultant neutrino mixing matrix
contains both Dirac and Majorana phases of CP violation, is also
discussed.
\end{abstract}

\pacs{PACS number(s): 14.60.Lm, 14.60.Pq, 95.85.Ry}

\newpage

\section{Introduction}

The solar \cite{SNO}, atmospheric \cite{SK}, reactor \cite{KM} and
accelerator \cite{K2K} neutrino experiments have provided us with
very convincing evidence that neutrinos are massive and lepton
flavors are mixed. Given the basis in which the flavor eigenstates
of charged leptons are identified with their mass eigenstates, the
phenomenon of neutrino mixing can simply be described by a
$3\times 3$ unitary matrix $V$ which transforms the neutrino mass
eigenstates $(\nu^{}_1, \nu^{}_2, \nu^{}_3)$ into the neutrino
flavor eigenstates $(\nu^{}_e, \nu^{}_\mu, \nu^{}_\tau)$. We
assume massive neutrinos to be Majorana particles and parametrize
$V$ as
\begin{equation}
V = \left( \matrix{ c^{}_{12}c^{}_{13} & s^{}_{12}c ^{}_{13} &
s^{}_{13} e^{-i\delta} \cr -s^{}_{12}c^{}_{23}
-c^{}_{12}s^{}_{23}s^{}_{13} e^{i\delta} & c^{}_{12}c^{}_{23}
-s^{}_{12}s^{}_{23}s^{}_{13} e^{i\delta} & s^{}_{23}c^{}_{13} \cr
s^{}_{12}s^{}_{23} -c^{}_{12}c^{}_{23}s^{}_{13} e^{i\delta} &
-c^{}_{12}s^{}_{23} -s^{}_{12}c^{}_{23}s^{}_{13} e^{i\delta} &
c^{}_{23}c^{}_{13} } \right) \left ( \matrix{e^{i\rho } & 0 & 0
\cr 0 & e^{i\sigma} & 0 \cr 0 & 0 & 1 \cr} \right ) \; ,
\end{equation}
where $c^{}_{ij} \equiv \cos\theta_{ij}$ and $s^{}_{ij} \equiv
\sin\theta_{ij}$ (for $ij=12,23$ and $13$). A global analysis of
current experimental data \cite{Vissani} yields $30^\circ <
\theta^{}_{12} < 38^\circ$, $36^\circ < \theta^{}_{23} < 54^\circ$
and $\theta^{}_{13} < 10^\circ$ at the $99\%$ confidence level,
but three CP-violating phases of $V$ (i.e., the Dirac phase
$\delta$ and the Majorana phases $\rho$ and $\sigma$) are entirely
unrestricted. In order to interpret the largeness of two neutrino
mixing angles together with the smallness of three neutrino
masses, many theoretical and phenomenological models of lepton
mass matrices have been proposed in the literature \cite{Review}.
Among them, the scenarios based on possible flavor symmetries are
particularly simple, suggestive and predictive.

In this paper, we focus our interest on the neutrino mass model
proposed recently by Friedberg and Lee (FL) \cite{Lee}. The
neutrino mass operator in the FL model is simply given by
\begin{eqnarray}
{\cal L}^{}_{\nu- \rm mass} & = & a \left (\overline{\nu}^{}_\tau
- \overline{\nu}^{}_\mu \right ) \left (\nu^{}_\tau - \nu^{}_\mu
\right ) + b \left (\overline{\nu}^{}_\mu - \overline{\nu}^{}_e
\right ) \left (\nu^{}_\mu - \nu^{}_e \right ) + c \left
(\overline{\nu}^{}_e - \overline{\nu}^{}_\tau \right ) \left
(\nu^{}_e - \nu^{}_\tau \right ) ~~~~~~~
\nonumber \\
& & + m^{}_0 \left (\overline{\nu}^{}_e \nu^{}_e +
\overline{\nu}^{}_\mu \nu^{}_\mu + \overline{\nu}^{}_\tau
\nu^{}_\tau \right ) \; ,
\end{eqnarray}
where $a$, $b$, $c$ and $m^{}_0$ are all assumed to be {\it real}.
A salient feature of ${\cal L}^{}_{\nu- \rm mass}$ is its partial
gauge-like symmetry; i.e., its $a$, $b$ and $c$ terms are
invariant under the transformation $\nu^{}_\alpha \rightarrow
\nu^{}_\alpha + z$ (for $\alpha = e, \mu, \tau$) with $z$ being a
space-time independent constant element of the Grassmann algebra
\cite{Lee}. Eq. (2) means that the neutrino mass matrix $M^{}_\nu$
takes the form
\begin{equation}
M^{}_\nu \; = \; m^{}_0 \left ( \matrix{1 & 0 & 0 \cr 0 & 1 & 0
\cr 0 & 0 & 1 \cr} \right ) ~ + ~ \left ( \matrix{ b+c & -b & -c
\cr -b & a+b & -a \cr -c & -a & a+c \cr} \right ) \; .
\end{equation}
Diagonalizing $M^{}_\nu$ by the transformation $V^{\dagger}_{\rm
FL} M^{}_\nu V^{*}_{\rm FL} = {\rm Diag} \{ m^{}_1, m^{}_2, m^{}_3
\}$, in which $m^{}_i$ (for $i=1,2,3$) stand for the neutrino
masses, one may obtain the neutrino mixing matrix
\begin{equation}
V^{}_{\rm FL} = \left ( \matrix{ \displaystyle \frac{2}{\sqrt{6}}
& \displaystyle \frac{1}{\sqrt{3}} & 0 \cr\cr \displaystyle
-\frac{1}{\sqrt{6}} & \displaystyle \frac{1}{\sqrt{3}} &
\displaystyle \frac{1}{\sqrt{2}} \cr\cr \displaystyle
-\frac{1}{\sqrt{6}} & \displaystyle \frac{1}{\sqrt{3}} &
\displaystyle -\frac{1}{\sqrt{2}} \cr} \right ) \left ( \matrix{
\displaystyle \cos\frac{\theta}{2} & 0 & \displaystyle
\sin\frac{\theta}{2} \cr\cr 0 & 1 & 0 \cr\cr \displaystyle
-\sin\frac{\theta}{2} & 0 & \displaystyle \cos\frac{\theta}{2}
\cr} \right ) \; ,
\end{equation}
where $\theta$ is given by $\tan\theta = \sqrt{3} \left (b-c
\right )/\left [ \left (b + c\right ) - 2a \right ]$. This
interesting result leads us to the following observations:
\begin{itemize}
\item       If $\theta =0^\circ$ holds, $V^{}_{\rm FL}$ will
reproduce the exact tri-bimaximal neutrino mixing pattern (with
$\tan\theta^{}_{12} = 1/\sqrt{2}$ or $\theta^{}_{12} \approx
35.3^\circ$, $\theta^{}_{23} = 45^\circ$ and $\theta^{}_{13} =
0^\circ$) \cite{TB}. The latter, which can be understood as a
geometric representation of the neutrino mixing matrix \cite{TD},
is in good agreement with current experimental data. Non-vanishing
but small $\theta$ predicts $\sin\theta^{}_{13} = ( 2/\sqrt{6})
\sin \left (\theta/2 \right )$, implying $\theta \lesssim
24.6^\circ$ for $\theta^{}_{13} < 10^\circ$. On the other hand,
$\theta^{}_{23}$ will mildly deviate from its best-fit value
$\theta^{}_{23} = 45^\circ$ if $\theta$ (or $\theta^{}_{13}$)
takes non-zero values.

\item       The limit $\theta = 0^\circ$ results from $b = c$.
When $b=c$ holds, it is easy to check that the neutrino mass
operator ${\cal L}^{}_{\nu- \rm mass}$ has the exact $\mu$-$\tau$
symmetry (i.e., ${\cal L}^{}_{\nu- \rm mass}$ is invariant under
the exchange of $\mu$ and $\tau$ indices). In other words, the
tri-bimaximal neutrino mixing is a natural consequence of the
$\mu$-$\tau$ symmetry of $M^{}_\nu$ in the FL model. Then
$\theta^{}_{13} \neq 0^\circ$ and $\theta^{}_{23} \neq 45^\circ$
measure the strength of $\mu$-$\tau$ symmetry breaking, as many
authors have discussed in other neutrino mass models
\cite{Symmetry}.
\end{itemize}
In addition, one may consider to remove one degree of freedom from
${\cal L}^{}_{\nu- \rm mass}$ or $M^{}_\nu$ (for instance, by
setting $c =0$ \cite{Lee}).

We aim to generalize the FL model to accommodate CP and T
violation for massive Majorana neutrinos
\footnote{To include T violation into the model, Friedberg and Lee
\cite{Lee} have inserted the phase factors $e^{\pm i \eta}$ into
Eq. (2) by replacing the term $c \left (\overline{\nu}^{}_e -
\overline{\nu}^{}_\tau \right ) \left (\nu^{}_e - \nu^{}_\tau
\right )$ with the term $c \left (e^{-i\eta} \overline{\nu}^{}_e -
\overline{\nu}^{}_\tau \right ) \left (e^{+i\eta} \nu^{}_e -
\nu^{}_\tau \right )$. The resultant neutrino mass matrix is no
longer symmetric, hence it definitely describes Dirac neutrinos
instead of Majorana neutrinos.}.
The effective Majorana neutrino mass term can be written as
\begin{equation}
{\cal L}^{\prime}_{\nu- \rm mass} \; = \; \frac{1}{2} ~
\overline{\left (\nu^{}_e, ~ \nu^{}_\mu, ~ \nu^{}_\tau \right
)^{}_{\rm L}} ~ M^{}_{\nu} \left ( \matrix{\nu^{\rm c}_e \cr
\nu^{\rm c}_\mu \cr \nu^{\rm c}_\tau \cr} \right )^{}_{\rm R} +
{\rm h.c.} \; ,
\end{equation}
where $\nu^{\rm c}_\alpha \equiv C \bar{\nu}^T_\alpha$ (for
$\alpha = e, \mu, \tau$), and $M^{}_\nu$ is of the same form as
that given in Eq. (3). Now the parameters of $M^{}_\nu$ (i.e.,
$a$, $b$, $c$ and $m^{}_0$) are all complex. Then we are able to
derive both the Dirac phase ($\delta$) and the Majorana phases
($\rho$ and $\sigma$) for the neutrino mixing matrix $V$. Two
special cases are particularly interesting:
\begin{itemize}
\item       Scenario (A): $a$ and $m^{}_0$ are real, and $b = c^*$
are complex. We find that the $\mu$-$\tau$ symmetry of $M^{}_\nu$
is softly broken in this case, leading to the elegant predictions
$\theta^{}_{13} \neq 0^\circ$, $\theta^{}_{23} =45^\circ$ and
$\delta = 90^\circ$. Two Majorana phases $\rho$ and $\sigma$ keep
vanishing.

\item       Scenario (B): $a$, $b$ and $c$ are real, but $m^{}_0$
is complex. We find that the results of $\theta^{}_{12}$,
$\theta^{}_{23}$ and $\theta^{}_{13}$ obtained from $V^{}_{\rm
FL}$ keep unchanged in this case, but some nontrivial values of
the Majorana phases $\rho$ and $\sigma$ can now be generated. The
Dirac phase $\delta$ remains vanishing.
\end{itemize}

The remaining part of this paper is organized as follows. More
detailed discussions about scenarios (A) and (B) will be presented
in section II. Section III is devoted to a generic analysis of the
generalized FL model with no special assumptions. Finally, we make
some concluding remarks in section IV.

\section{Two simple scenarios}

First of all, let us consider two special but interesting
scenarios of the generalized FL model and explore their respective
consequences on three neutrino mixing angles and three
CP-violating phases. They are quite instructive in phenomenology
and may easily be tested by a variety of long-baseline neutrino
oscillation experiments in the near future.

\subsection{Scenario (A)}

In this scenario, $a$ and $m^{}_0$ are real, $b$ and $c$ are
complex, and $b = c^*$ holds. The corresponding neutrino mass
matrix $M^{}_\nu$ reads
\begin{equation}
M^{}_\nu \; = \; m^{}_0 \left ( \matrix{1 & 0 & 0 \cr 0 & ~ 1 ~ &
0 \cr 0 & 0 & 1 \cr} \right ) ~ + ~ \left ( \matrix{ 2{\rm Re}
\left (b \right ) & -b & -b^* \cr -b & a + b & -a \cr -b^* & -a &
a + b^* \cr} \right ) \; .
\end{equation}
We find that $M^{}_\nu$ can be diagonalized by the transformation
$V^{\dagger} M^{}_\nu V^{*} = {\rm Diag} \{ m^{}_1, m^{}_2, m^{}_3
\}$, where
\begin{equation}
V = \left ( \matrix{ \displaystyle \frac{2}{\sqrt{6}} &
\displaystyle \frac{1}{\sqrt{3}} & 0 \cr\cr \displaystyle
-\frac{1}{\sqrt{6}} & \displaystyle \frac{1}{\sqrt{3}} &
\displaystyle \frac{1}{\sqrt{2}} \cr\cr \displaystyle
-\frac{1}{\sqrt{6}} & \displaystyle \frac{1}{\sqrt{3}} &
\displaystyle -\frac{1}{\sqrt{2}} \cr} \right ) \left ( \matrix{
\displaystyle \cos\frac{\theta}{2} & 0 & \displaystyle
\sin\frac{\theta}{2} e^{-i\delta} \cr\cr 0 & 1 & 0 \cr\cr
\displaystyle -\sin\frac{\theta}{2} e^{i\delta} & 0 & \displaystyle
\cos\frac{\theta}{2} \cr} \right ) \; ,
\end{equation}
and $\delta$ is just the Dirac phase of CP violation defined in
the standard parametrization of $V$. A straightforward calculation
yields $\delta =  90^\circ$,
\begin{equation}
\tan\theta \; =\; \frac{\sqrt{3} ~ {\rm Im} \left (b \right )}
{m^{}_0 + a + 2 {\rm Re} \left (b \right )} \; ,
\end{equation}
together with three neutrino masses
\begin{eqnarray}
m^{}_1 & = &  \sqrt{\left [m^{}_0 + a + 2{\rm Re} \left (b \right
) \right ]^2 + 3 \left [ {\rm Im} \left (b \right ) \right ]^2} ~
- a + {\rm Re} \left (b \right ) \; ,
\nonumber \\
m^{}_2 & = & m^{}_0 \; ,
\nonumber \\
m^{}_3 & = & \sqrt{\left [m^{}_0 + a + 2{\rm Re} \left (b \right )
\right ]^2 + 3 \left [ {\rm Im} \left (b \right ) \right ]^2} ~ +
a - {\rm Re} \left (b \right ) \; .
\end{eqnarray}
Note that the special result of $\delta$ is a natural consequence
of the purely imaginary term $b - c = 2i {\rm Im} \left (b \right
)$ in this scenario. The difference between $b$ and $c$ can be
referred to as the {\it soft} $\mu$-$\tau$ symmetry breaking,
because $|b| = |c|$ holds. The explicit expression of $V$ is
\begin{equation}
V = \left ( \matrix{ \displaystyle \frac{2}{\sqrt{6}}
\cos\frac{\theta}{2} & \displaystyle \frac{1}{\sqrt{3}} &
\displaystyle -i \frac{2}{\sqrt{6}} \sin\frac{\theta}{2} \cr\cr
\displaystyle -\frac{1}{\sqrt{6}} \cos\frac{\theta}{2} - i
\frac{1}{\sqrt{2}} \sin\frac{\theta}{2} & \displaystyle
\frac{1}{\sqrt{3}} & \displaystyle \frac{1}{\sqrt{2}}
\cos\frac{\theta}{2} + i \frac{1}{\sqrt{6}} \sin\frac{\theta}{2}
\cr\cr \displaystyle -\frac{1}{\sqrt{6}} \cos\frac{\theta}{2} + i
\frac{1}{\sqrt{2}} \sin\frac{\theta}{2} & \displaystyle
\frac{1}{\sqrt{3}} & \displaystyle -\frac{1}{\sqrt{2}}
\cos\frac{\theta}{2} + i \frac{1}{\sqrt{6}} \sin\frac{\theta}{2}
\cr} \right ) \; .
\end{equation}
We observe that the only difference between $V$ in Eq. (10) and
$V^{}_{\rm FL}$ in Eq. (4) is the introduction of a special
CP-violating phase (see also Ref. \cite{Yasue} for a discussion
about the maximal leptonic CP violation with $\delta = 90^\circ$).
This CP-violating phase, which is attributed to the soft breaking
of $\mu$-$\tau$ symmetry, can change the prediction of $V^{}_{\rm
FL}$ for $\theta^{}_{23}$. Comparing Eq. (10) with Eq. (1), we
immediately obtain
\footnote{It is trivial to redefine the phases of charged-lepton
fields, such that the location of CP-violating phases in Eq. (10)
is the same as that in Eq. (1).}
\begin{eqnarray}
\sin\theta^{}_{12} & = & \frac{1}{\sqrt{2 + \cos\theta}} \; ,
\nonumber \\
\sin\theta^{}_{23} & = & \frac{1}{\sqrt{2}} \; , \nonumber \\
\sin\theta^{}_{13} & = & \frac{2}{\sqrt{6}} \sin\frac{\theta}{2}
\; ,
\end{eqnarray}
together with $\delta =  90^\circ$ and $\rho = \sigma = 0^\circ$.
Without loss of generality, we have restricted $\theta$ to the
first quadrant. The leptonic Jarlskog parameter $\cal J$ \cite{J},
which is a rephasing-invariant measure of CP violation in neutrino
oscillations, reads ${\cal J} = \sin\theta /(6\sqrt{3})$. One can
see that the soft breaking of $\mu$-$\tau$ symmetry leads to both
$\theta^{}_{13} \neq 0^\circ$ and ${\cal J} \neq 0$, but it does
not affect the favorable result $\theta^{}_{23} = 45^\circ$ given
by the tri-bimaximal neutrino mixing pattern. On the other hand,
$\sin\theta^{}_{12} \approx 1/\sqrt{3}$ is an excellent
approximation, since $\theta$ must be small to maintain the
smallness of $\theta^{}_{13}$. In view of $\theta^{}_{13} <
10^\circ$, we obtain $\theta \lesssim 24.6^\circ$ and ${\cal J}
\lesssim 0.04$. It is possible to measure ${\cal J} \sim {\cal
O}(10^{-2})$ in the future long-baseline neutrino oscillation
experiments.

Note that the neutrino masses rely on four real model parameters
$m^{}_0$, $a$, ${\rm Re}(b)$ and ${\rm Im}(b)$. Thus it is easy to
fit the neutrino mass-squared differences $\Delta m^2_{21} = (7.2
\cdots 8.9) \times 10^{-5} ~ {\rm eV}^2$ and $\Delta m^2_{32} =
\pm (2.1 \cdots 3.1) \times 10^{-3} ~ {\rm eV}^2$ \cite{Vissani}.
Such a fit should not involve any fine-tuning, because (a) the
number of free parameters is larger than the number of constraint
conditions and (b) three neutrino masses have very weak
correlation with three mixing angles. A detailed numerical
analysis shows that only the normal mass hierarchy ($m^{}_1 <
m^{}_2 < m^{}_3$) is allowed in this scenario. FIG. 1 illustrates
the parameter space of $m^{}_0$, $a$, ${\rm Re}(b)$ and ${\rm
Im}(b)$, where $m^{}_0 \lesssim 0.2$ eV has typically been taken
as a generous upper bound on the absolute neutrino mass
\cite{WMAP}. Because of $m^{}_0 = m^{}_2$, it is straightforward
to get $m^{}_0 > \sqrt{\Delta m^2_{21}} \approx 0.009 ~ {\rm eV}$,
as shown in FIG. 1. The small $\mu$-$\tau$ symmetry breaking (or
small $\theta$) requires the small magnitude of ${\rm Im}(b)$.
Thus Eq. (10) allows us to get ${\rm Re}(b) \approx (m^{}_1 -
m^{}_2)/3$ in the ${\rm Im}(b) \rightarrow 0$ limit, implying that
${\rm Re}(b)$ is negative and its magnitude is very small.
Furthermore, $a \approx (m^{}_3 - m^{}_2)/2$ approximately holds
in the ${\rm Im}(b) \rightarrow 0$ limit, implying that the
maximal value of $a$ is roughly half of $m^{}_3 \sim \sqrt{|\Delta
m^2_{32}|}$ (i.e., $a \lesssim 0.025 ~ {\rm eV}$) when three
neutrino masses develop a strong normal hierarchy.

\subsection{Scenario (B)}

In this simple scenario, only $m^{}_0$ is assumed to be complex.
While the neutrino mass matrix $M^{}_\nu$ takes the same form as
that given in Eq. (3), its diagonalization ($V^\dagger M^{}_\nu
V^* = {\rm Diag} \{ m^{}_1, m^{}_2, m^{}_3 \}$) requires the
following transformation matrix
\begin{equation}
V = \left ( \matrix{ \displaystyle \frac{2}{\sqrt{6}} &
\displaystyle \frac{1}{\sqrt{3}} & 0 \cr\cr \displaystyle
-\frac{1}{\sqrt{6}} & \displaystyle \frac{1}{\sqrt{3}} &
\displaystyle \frac{1}{\sqrt{2}} \cr\cr \displaystyle
-\frac{1}{\sqrt{6}} & \displaystyle \frac{1}{\sqrt{3}} &
\displaystyle -\frac{1}{\sqrt{2}} \cr} \right ) \left ( \matrix{
\displaystyle \cos\frac{\theta}{2} & 0 & \displaystyle
\sin\frac{\theta}{2} \cr\cr 0 & 1 & 0 \cr\cr \displaystyle
-\sin\frac{\theta}{2} & 0 & \displaystyle \cos\frac{\theta}{2} \cr}
\right ) \left ( \matrix{ e^{i\phi^{}_1} & 0 & 0 \cr\cr 0 &
e^{i\phi^{}_2} & 0 \cr\cr 0 & 0 & e^{i\phi^{}_3} \cr} \right ) \; ,
\end{equation}
where $\phi^{}_i$ (for $i=1,2,3$) originate from the imaginary
part of $m^{}_0$. Comparing between Eqs. (1) and (12), one can see
that the Majorana phases of CP violation in the standard
parametrization of $V$ are given by $\rho = \phi^{}_1 - \phi^{}_3$
and $\sigma = \phi^{}_2 - \phi^{}_3$ (i.e., the phase factor
$e^{i\phi^{}_3}$ in Eq. (12) is finally rotated away by redefining
the phases of three charged-lepton fields). After a
straightforward calculation, we obtain
\begin{equation}
\tan\theta \; =\; \frac{\sqrt{3} \left (b - c \right )}{\left (b +
c \right ) - 2a} \; ,
\end{equation}
and
\begin{eqnarray}
m^{}_1 & = & \sqrt{\left [m^{}_{-} + {\rm Re} \left (m^{}_0 \right ) \right ]^2
+ \left [ {\rm Im} \left ( m^{}_0 \right ) \right ]^2} \;\; , \nonumber \\
m^{}_2 & = & \left |m^{}_0 \right | \; , \nonumber \\
m^{}_3 & = & \sqrt{\left [m^{}_{+} + {\rm Re} \left (m^{}_0 \right
) \right ]^2 + \left [ {\rm Im} \left ( m^{}_0 \right ) \right
]^2} \;\; ,
\end{eqnarray}
where
\begin{equation}
m^{}_{\pm} \; = \; \left ( a + b + c \right ) \pm \sqrt{a^2 + b^2
+ c^2 - ab - ac - bc} \;\; .
\end{equation}
The final result for the neutrino mixing matrix $V$ is
\begin{equation}
V = \left ( \matrix{ \displaystyle \frac{2}{\sqrt{6}}
\cos\frac{\theta}{2} & \displaystyle \frac{1}{\sqrt{3}} &
\displaystyle \frac{2}{\sqrt{6}} \sin\frac{\theta}{2} \cr\cr
\displaystyle -\frac{1}{\sqrt{6}} \cos\frac{\theta}{2} -
\frac{1}{\sqrt{2}} \sin\frac{\theta}{2} & \displaystyle
\frac{1}{\sqrt{3}} & \displaystyle \frac{1}{\sqrt{2}}
\cos\frac{\theta}{2} - \frac{1}{\sqrt{6}} \sin\frac{\theta}{2}
\cr\cr \displaystyle -\frac{1}{\sqrt{6}} \cos\frac{\theta}{2} +
\frac{1}{\sqrt{2}} \sin\frac{\theta}{2} & \displaystyle
\frac{1}{\sqrt{3}} & \displaystyle -\frac{1}{\sqrt{2}}
\cos\frac{\theta}{2} - \frac{1}{\sqrt{6}} \sin\frac{\theta}{2}
\cr} \right ) \left ( \matrix{ e^{i\rho} & 0 & 0 \cr\cr 0 &
e^{i\sigma} & 0 \cr\cr 0 & 0 & 1 \cr} \right ) \; ,
\end{equation}
where $\rho$ and $\sigma$ are given by
\begin{eqnarray}
\tan 2 \rho & = & \frac{\left ( m^{}_+ - m^{}_- \right ) {\rm Im}
\left ( m^{}_0 \right )}{|m^{}_0|^2 + m^{}_+ m^{}_- + \left (
m^{}_+ + m^{}_- \right ) {\rm Re} \left ( m^{}_0 \right )} \; ,
\nonumber \\
\tan 2 \sigma & = & \frac{ m^{}_+ {\rm Im} \left ( m^{}_0 \right
)}{|m^{}_0|^2 + m^{}_+ {\rm Re} \left ( m^{}_0 \right )} \; .
\end{eqnarray}
One can see that the only difference between $V$ in Eq. (16) and
$V^{}_{\rm FL}$ in Eq. (4) is the introduction of two Majorana
phases of CP violation. Although $\rho$ and $\sigma$ have nothing
to do with the behaviors of neutrino oscillations, they may
significantly affect the neutrinoless double-beta decay
\cite{Xing02}. Comparing Eq. (16) with Eq. (1), we arrive at
\begin{eqnarray}
\sin\theta^{}_{12} & = & \frac{1}{\sqrt{2 + \cos\theta}} \; ,
\nonumber \\
\sin\theta^{}_{23} & = & \frac{\sqrt{2 + \cos\theta - \sqrt{3}
\sin\theta}}{\sqrt{2 \left (2 + \cos\theta \right )}} \; ,
\nonumber \\
\sin\theta^{}_{13} & = & \frac{2}{\sqrt{6}} \sin\frac{\theta}{2}
\; ,
\end{eqnarray}
together with $\delta = 0^\circ$ for the Dirac phase of CP
violation. Again $\theta$ has been restricted to the first
quadrant. The results for $\theta^{}_{12}$ and $\theta^{}_{13}$ in
this scenario are the same as those obtained in scenario (A), but
the Jarlskog parameter $\cal J$ is now vanishing. Because of the
$\mu$-$\tau$ symmetry breaking, $\theta^{}_{23}$ may somehow
deviate from the favorable value $\theta^{}_{23} = 45^\circ$.
Given $\theta \lesssim 24.6^\circ$ corresponding to
$\theta^{}_{13} < 10^\circ$, $\theta^{}_{23}$ is allowed to vary
in the range $37.8^\circ \lesssim \theta^{}_{23} \leq 45^\circ$
\cite{US}.

Note that the neutrino masses depend on five real model parameters
$a$, $b$, $c$, ${\rm Re}(m^{}_0)$ and ${\rm Im}(m^{}_0)$. Hence
there is sufficient freedom to fit two observed neutrino
mass-squared differences $\Delta m^2_{21}$ and $\Delta m^2_{32}$.
A detailed numerical analysis shows that both the normal mass
hierarchy ($m^{}_1 < m^{}_2 < m^{}_3$) and the inverted mass
hierarchy ($m^{}_3 < m^{}_1 < m^{}_2$) are allowed in scenario
(B). FIGs. 2 and 3 illustrate the parameter space of $a$, $b$,
$c$, ${\rm Re}(m^{}_0)$ and ${\rm Im}(m^{}_0)$, where $|m^{}_0|
\lesssim 0.2$ eV has been taken \cite{WMAP}. Taking account of the
analytical reciprocity between $b$ and $c$ (i.e., an exchange of
$b$ and $c$ in Eq. (15) does not affect the result of $m^{}_i$
obtained in Eq. (14)), we plot their allowed regions in the same
figure without any confusion. Note that $m^{}_0$ cannot be purely
imaginary in both normal and inverted mass hierarchies, otherwise
we would be left with $\Delta m^2_{21} < 0$, which is in
disagreement with current experimental data. Moreover, the lower
bound of $|{\rm Re}(m^{}_0)|$ is obviously restricted by the
magnitude of $m^{}_2$ (i.e., $m^{}_2 \sim \sqrt{\Delta m^2_{21}}
\approx 0.009 ~ {\rm eV}$ in the normal mass hierarchy or $m^{}_2
\sim \sqrt{|\Delta m^2_{32}|} \approx 0.05 ~ {\rm eV}$ in the
inverted mass hierarchy). In view of $(m^{}_+ - m^{}_-) \propto
(m^2_3 - m^2_1)$ from Eq. (14), we find $m^{}+ \neq m^{}_-$. This
inequality implies that $a=b=c$ is not allowed, as one can see
from Eq. (15) or FIGs. 2 and 3.
The disconnected regions in the plots of $a$ versus $b$ (or $c$)
are ascribed to the ambiguity induced by the sign of ${\rm
Re}(m^{}_0)$. FIG. 4 shows the allowed regions of $\rho$ and
$\sigma$. We see that both of them are less restricted, as a
consequence of the large freedom associated with the imaginary
part of $m^{}_0$.

\section{Generic analysis}

Now let us assume all the parameters of $M^{}_\nu$ in Eq. (3) to
be complex and calculate its mass eigenvalues and flavor mixing
parameters. Using the tri-bimaximal mixing matrix \cite{TB}
\begin{equation}
V^{}_0 = \left ( \matrix{ \displaystyle \frac{2}{\sqrt{6}} &
\displaystyle \frac{1}{\sqrt{3}} & 0 \cr\cr \displaystyle
-\frac{1}{\sqrt{6}} & \displaystyle \frac{1}{\sqrt{3}} &
\displaystyle \frac{1}{\sqrt{2}} \cr\cr \displaystyle
-\frac{1}{\sqrt{6}} & \displaystyle \frac{1}{\sqrt{3}} &
\displaystyle -\frac{1}{\sqrt{2}} \cr} \right ) \; ,
\end{equation}
we transform $M^{}_\nu$ into the following form:
\begin{equation}
V^\dagger_0 M^{}_\nu V^{*}_0 \; = \; \left(\matrix{\displaystyle
\frac{3}{2} \left (b + c \right ) + m^{}_0 & 0 & \displaystyle
\frac{\sqrt{3}}{2} \left (c - b \right ) \cr 0 & m^{}_0 & 0 \cr
\displaystyle \frac{\sqrt{3}}{2} \left (c - b \right ) & 0 &
\displaystyle 2a + \frac{1}{2} \left (b + c \right ) + m^{}_0 \cr
}\right) \; .
\end{equation}
This complex mass matrix can be diagonalized by the transformation
$V^\dagger_1 (V^\dagger_0 M^{}_\nu V^{*}_0 ) V^{*}_1 = {\rm
Diag}\{m^{}_1, m^{}_2, m^{}_3 \}$, where
\begin{equation}
V^{}_1 \; = \; \left ( \matrix{ \displaystyle \cos\frac{\theta}{2}
& 0 & \displaystyle \sin\frac{\theta}{2} e^{-i\delta} \cr\cr 0 & 1
& 0 \cr\cr \displaystyle -\sin\frac{\theta}{2} e^{i\delta} & 0 &
\displaystyle \cos\frac{\theta}{2} \cr} \right ) \left ( \matrix{
e^{i\phi^{}_1} & 0 & 0 \cr\cr 0 & e^{i\phi^{}_2} & 0 \cr\cr 0 & 0
& e^{i\phi^{}_3} \cr} \right ) \; .
\end{equation}
The neutrino mixing matrix $V$ turns out to be $V = V^{}_0
V^{}_1$, in which $\delta$ is just the Dirac phase of CP
violation. Of course, $\rho = \phi^{}_1 - \phi^{}_3$ and $\sigma =
\phi^{}_2 - \phi^{}_3$ are the Majorana phases of CP violation in
the standard parametrization of $V$.

To be explicit, we express $\theta$ and $\delta$ in terms of the
parameters of $M^{}_\nu$. The results are
\begin{eqnarray}
\tan \theta & = & \frac{\sqrt{X^2 + Y^2}}{Z} \; ,
\nonumber \\
\tan\delta & = & \frac{X}{Y} \; ,
\end{eqnarray}
where
\begin{eqnarray}
X & = & \sqrt{3}{\rm Im}\left (ac^*-ab^* - bc^* \right ) \; ,
\nonumber \\
Y & = & \sqrt{3} \left [|c|^2 - |b|^2 + {\rm Re} \left (ac^* -
ab^* - m^{}_0b^* + m^{}_0c^* \right ) \right ] \; ,
\nonumber \\
Z & = & 2|a|^2 - |b|^2 - |c|^2 + {\rm Re} \left ( ab^* + ac^* -
2bc^* + 2m^{}_0a^* - m^{}_0b^* - m^{}_0c^* \right ) \; .
\end{eqnarray}
Furthermore, three mass eigenvalues of $M^{}_\nu$ and two Majorana
phases of $V$ are found to be
\begin{eqnarray}
m^{}_1 & = & \left | T^{}_1 \cos^2\frac{\theta}{2} - T_2\sin\theta
e^{-i\delta} + T^{}_3\sin^2\frac{\theta}{2} e^{-i 2\delta} \right
| \; ,
\nonumber \\
m^{}_2 & = & \left |m^{}_0 \right | \; ,
\nonumber \\
m^{}_3 & = & \left | T^{}_3 \cos^2\frac{\theta}{2} +
T^{}_2\sin\theta e^{+i\delta} + T^{}_1\sin^2\frac{\theta}{2} e^{+i
2\delta} \right | \; ;
\end{eqnarray}
and
\begin{eqnarray}
\rho & = &  \frac{1}{2} \arg \left( \frac{\displaystyle T^{}_1
\cos^2\frac{\theta}{2} - T^{}_2\sin\theta e^{-i\delta} +
T^{}_3\sin^2\frac{\theta}{2} e^{-i 2\delta}}{\displaystyle T^{}_3
\cos^2\frac{\theta}{2} + T^{}_2\sin\theta e^{+i\delta} +
T^{}_1\sin^2\frac{\theta}{2} e^{+i 2\delta}} \right) \; ,
\nonumber \\
\sigma & = &  \frac{1}{2}\arg \left( \frac{m^{}_0}{\displaystyle
T^{}_3 \cos^2\frac{\theta}{2} + T^{}_2\sin\theta e^{+i\delta} +
T^{}_1\sin^2\frac{\theta}{2} e^{+i 2\delta}} \right) \; ,
\end{eqnarray}
where
\begin{eqnarray}
T^{}_{1} & = & \frac{3}{2} \left (b + c \right ) + m^{}_0 \; ,
\nonumber \\
T^{}_{2} & = & \frac{\sqrt{3}}{2} \left (c - b \right ) \; ,
\nonumber \\
T^{}_{3} & = & 2a + \frac{1}{2} \left (b+ c \right ) + m^{}_0 \; .
\end{eqnarray}
The final result of $V$ is
\begin{equation}
V = \left ( \matrix{ \displaystyle \frac{2}{\sqrt{6}}
\cos\frac{\theta}{2} & \displaystyle \frac{1}{\sqrt{3}} &
\displaystyle \frac{2}{\sqrt{6}} \sin\frac{\theta}{2} e^{-i\delta}
\cr\cr \displaystyle -\frac{1}{\sqrt{6}} \cos\frac{\theta}{2} -
\frac{1}{\sqrt{2}} \sin\frac{\theta}{2} e^{i\delta} &
\displaystyle \frac{1}{\sqrt{3}} & \displaystyle
\frac{1}{\sqrt{2}} \cos\frac{\theta}{2} - \frac{1}{\sqrt{6}}
\sin\frac{\theta}{2} e^{-i\delta} \cr\cr \displaystyle
-\frac{1}{\sqrt{6}} \cos\frac{\theta}{2} + \frac{1}{\sqrt{2}}
\sin\frac{\theta}{2} e^{i\delta} & \displaystyle
\frac{1}{\sqrt{3}} & \displaystyle -\frac{1}{\sqrt{2}}
\cos\frac{\theta}{2} - \frac{1}{\sqrt{6}} \sin\frac{\theta}{2}
e^{-i\delta} \cr} \right ) \left ( \matrix{ e^{i\rho} & 0 & 0
\cr\cr 0 & e^{i\sigma} & 0 \cr\cr 0 & 0 & 1 \cr} \right ) \; ,
\end{equation}
from which we obtain the Jarlskog parameter ${\cal J} =
\sin\theta\sin\delta/(6\sqrt{3})$. Comparing Eq. (27) with Eq.
(1), we arrive at
\begin{eqnarray}
\sin\theta^{}_{12} & = & \frac{1}{\sqrt{2 + \cos\theta}} \; ,
\nonumber \\
\sin\theta^{}_{23} & = & \frac{\sqrt{2 + \cos\theta -
\sqrt{3}\sin\theta\cos\delta}}{\sqrt{2 \left (2 + \cos\theta
\right )}} \; ,
\nonumber \\
\sin\theta^{}_{13} & = & \frac{2}{\sqrt{6}} \sin\frac{\theta}{2}
\; .
\end{eqnarray}
One can see that the predictions for $\theta^{}_{12}$ and
$\theta^{}_{13}$ are the same as those obtained in scenarios (A)
and (B). Thus they are two typical and general consequences of the
FL model. As for $\theta^{}_{23}$, the generic result in Eq. (28)
may easily reproduce the special result in Eq. (11) for scenario
(A) with $\delta =  90^\circ$ or that in Eq. (18) for scenario (B)
with $\delta = 0^\circ$. Given $\theta \lesssim 24.6^\circ$ (i.e.,
$\theta^{}_{13} < 10^\circ$), $\theta^{}_{23}$ is allowed to vary
in the range $37.8^\circ \leq \theta^{}_{23} \lesssim 52.2^\circ$
for arbitrary $\delta$.

As $m^{}_0$, $a$, $b$ and $c$ are all complex, we now have much
more freedom to fit two observed neutrino mass-squared differences
$\Delta m^2_{21}$ and $\Delta m^2_{32}$. Of course, both the
normal neutrino mass hierarchy ($m^{}_1 < m^{}_2 < m^{}_3$) and
the inverted neutrino mass hierarchy ($m^{}_3 < m^{}_1 < m^{}_2$)
are expected in this generic case. We shall not carry out a
numerical analysis of the parameter space, however, just because
it involves a lot of uncertainties and is not as suggestive as
that in scenario (A) or scenario (B).

\section{Concluding remarks}

We have pointed out that the nearly tri-bimaximal neutrino mixing
can be regarded as a natural consequence of the slight
$\mu$-$\tau$ symmetry breaking in the FL neutrino mass model.
Another straightforward consequence of the $\mu$-$\tau$ symmetry
breaking is CP violation. Assuming the Majorana nature of massive
neutrinos, we have generalized the FL model to introduce the
CP-violating effects. In addition to a generic analysis of the
generalized FL model, two simple but intriguing scenarios have
been proposed: scenario (A) involves the softly-broken
$\mu$-$\tau$ symmetry, leading to the elegant predictions
$\theta^{}_{13} \neq 0^\circ$, $\theta^{}_{23} = 45^\circ$,
$\delta =  90^\circ$ and vanishing Majorana phases of CP
violation; scenario (B) predicts $\theta^{}_{13} \neq 0^\circ$,
$\theta^{}_{23} \neq 45^\circ$, $\delta = 0^\circ$ and two
nontrivial Majorana phases of CP violation, on the other hand.
Both scenarios are compatible with current experimental data.

Although our discussions about the generalized FL model are
restricted to low-energy scales, it can certainly be extended to a
superhigh-energy scale (e.g., the GUT scale or the seesaw scale).
In this case, one should take into account the radiative
corrections to both neutrino masses and flavor mixing parameters
when they run from the high scale to the electroweak scale. A
particularly interesting point is that the Majorana phases $\rho$
and $\sigma$ can radiatively be generated from the Dirac phase
$\delta$ in scenario (A), while the Dirac phase $\delta$ can
radiatively be generated from the Majorana phases $\rho$ and
$\sigma$ in scenario (B) \cite{Luo1}. Thus the features of
leptonic CP violation can fully show up in both scenarios at
low-energy scales, if they are originally prescribed at a
superhigh-energy scale (see Ref. \cite{Luo2} for a detailed
analysis of the running behaviors of CP-violating phases in the
nearly tri-bimaximal neutrino mixing pattern, including the
minimal supergravity threshold effects).

We conclude that the $\mu$-$\tau$ symmetry and its slight breaking
are useful and suggestive for model building. We expect that a
stringent test of the generalized FL model, in particular its two
simple and instructive scenarios, can be achieved in the near
future from the reactor neutrino oscillation experiments (towards
measuring $\theta^{}_{13}$), the accelerator long-baseline
neutrino oscillation experiments (towards detecting both
$\theta^{}_{13}$ and $\delta$) and the neutrinoless double-beta
decay experiments (towards probing the Majorana nature of massive
neutrinos and constraining their Majorana phases of CP violation).

\acknowledgments{We are indebted to W.Q. Chao for bringing Refs.
\cite{Lee} and \cite{TD} to our particular attention. One of us
(Z.Z.X.) would also like to thank T.D. Lee for his encouragement
at the Workshop on Future China-US Cooperation in High Energy
Physics (June 2006, Beijing). This work is supported in part by
the National Natural Science Foundation of China.}

\newpage

\begin{figure}
\psfig{file=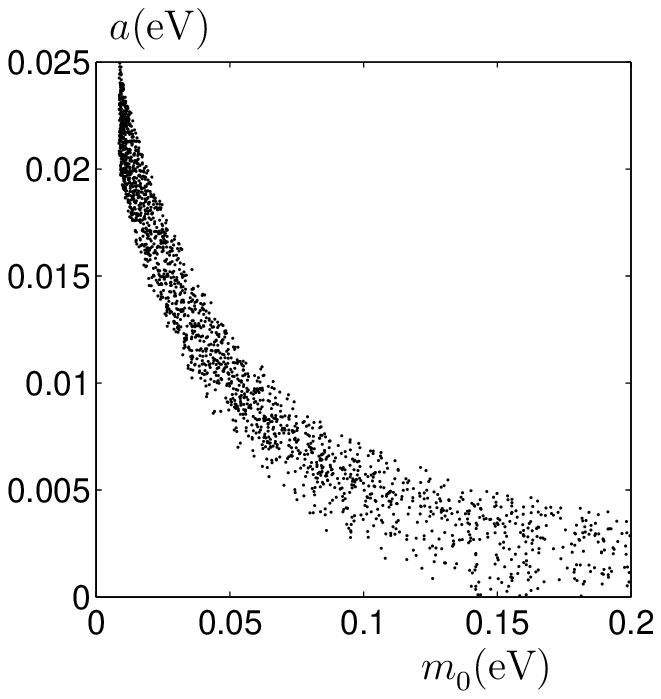, bbllx=2.2cm, bblly=6.0cm, bburx=12.2cm, bbury=16.0cm,%
width=12cm, height=12cm, angle=0, clip=0}
\end{figure}
\begin{figure}
\vspace{-3.5cm}
\psfig{file=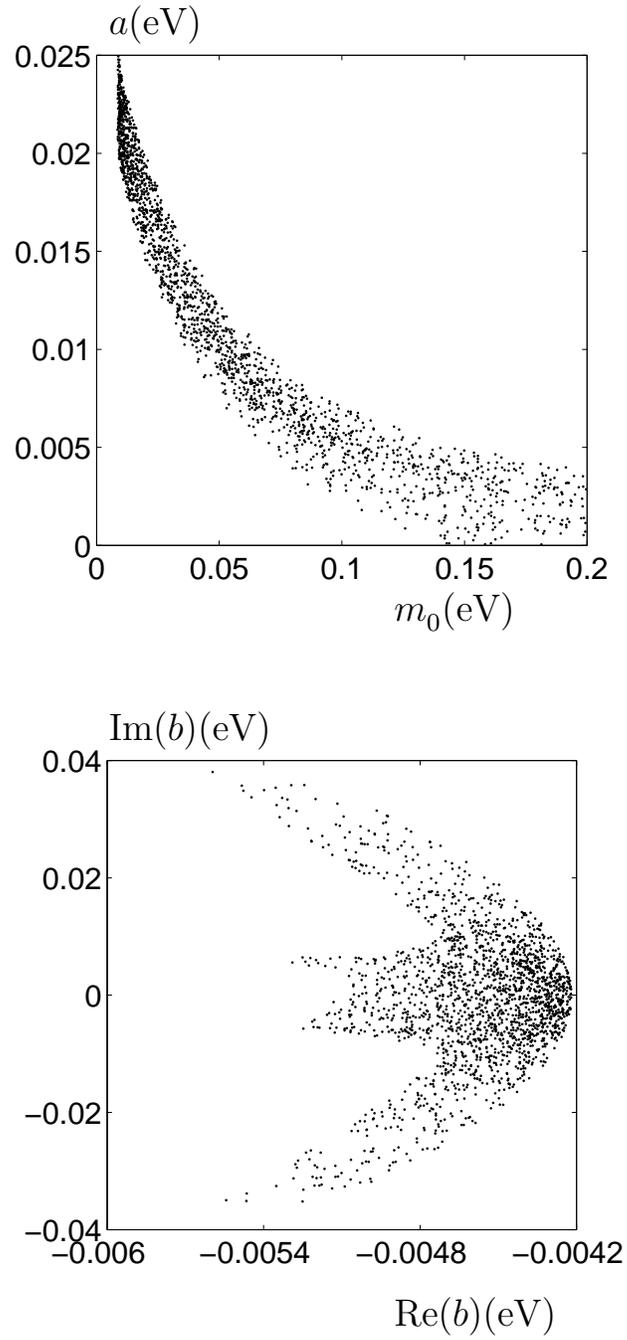, bbllx=2.2cm, bblly=6.0cm, bburx=12.2cm, bbury=16.0cm,%
width=12cm, height=12cm, angle=0, clip=0}\vspace{-2cm}\caption{
The parameter space of ($m^{}_0$, $a$) and (${\rm Re}(b)$, ${\rm
Im}(b)$) in \underline{\bf Scenario (A)}, where only the normal
neutrino mass hierarchy is allowed.}
\end{figure}

\newpage

\begin{figure}
\psfig{file=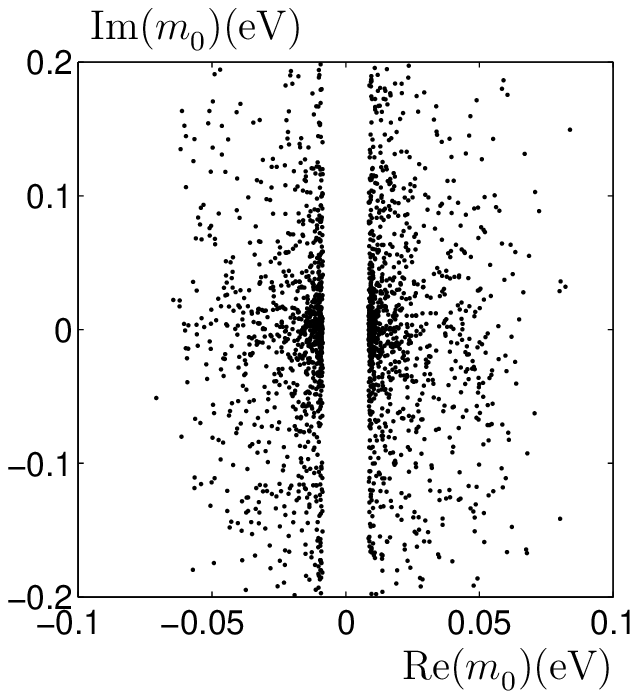, bbllx=2.2cm, bblly=6.0cm, bburx=12.2cm, bbury=16.0cm,%
width=12cm, height=12cm, angle=0, clip=0}
\end{figure}
\begin{figure}
\vspace{-3.5cm}
\psfig{file=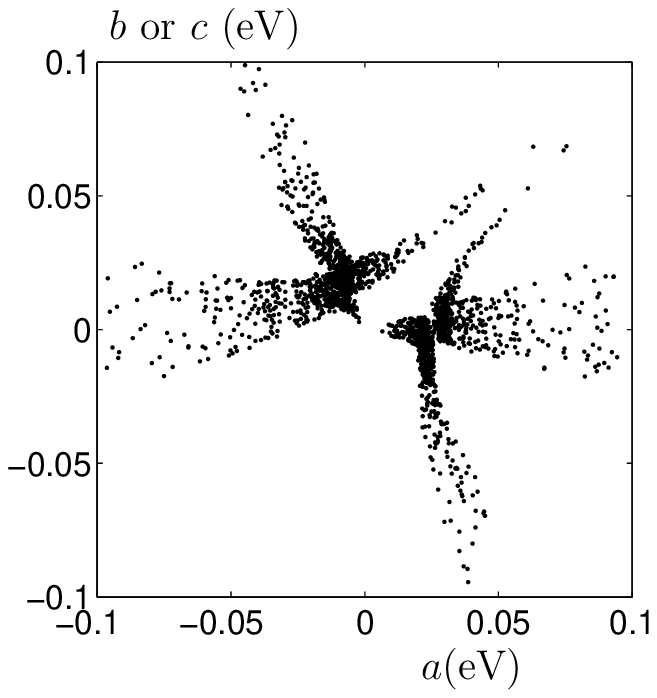, bbllx=2.2cm, bblly=6.0cm, bburx=12.2cm, bbury=16.0cm,%
width=12cm, height=12cm, angle=0, clip=0}\vspace{-2cm}\caption{The
parameter space of $\left({\rm Re}(m^{}_0), {\rm
Im}(m^{}_0)\right)$ and ($a$, $b$ or $c$) in \underline{\bf
Scenario (B)} with the {\it normal} neutrino mass hierarchy.}
\end{figure}

\newpage

\begin{figure}
\psfig{file=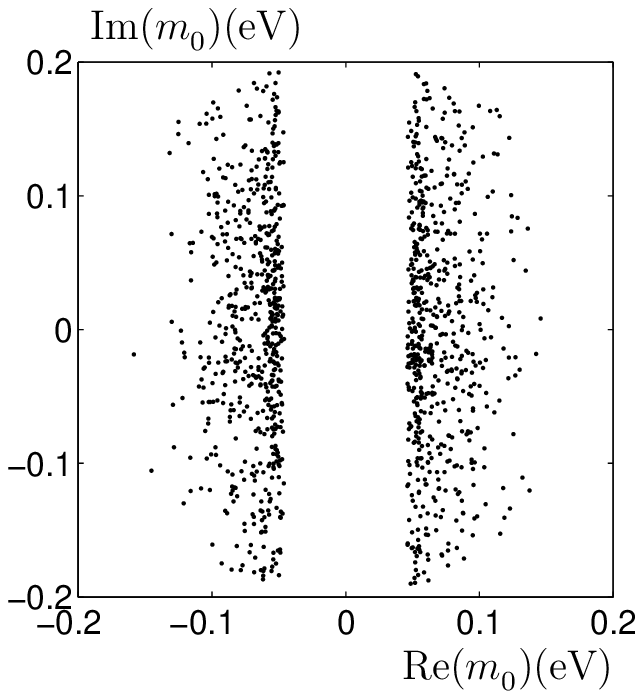, bbllx=2.2cm, bblly=6.0cm, bburx=12.2cm, bbury=16.0cm,%
width=12cm, height=12cm, angle=0, clip=0}
\end{figure}
\begin{figure}
\vspace{-3.5cm}
\psfig{file=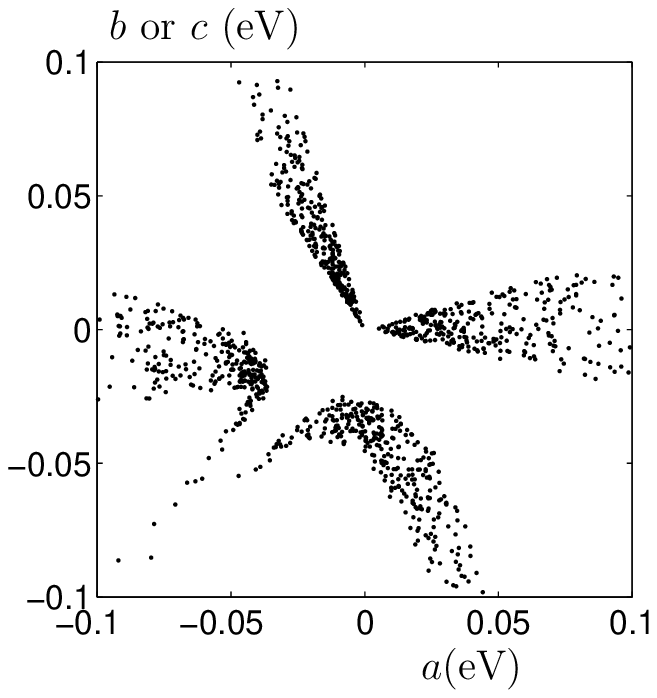, bbllx=2.2cm, bblly=6.0cm, bburx=12.2cm, bbury=16.0cm,%
width=12cm, height=12cm, angle=0, clip=0}\vspace{-2cm}\caption{The
parameter space of $\left({\rm Re}(m^{}_0), {\rm
Im}(m^{}_0)\right)$ and ($a$, $b$ or $c$) in \underline{\bf
Scenario (B)} with the {\it inverted} neutrino mass hierarchy.}
\end{figure}

\newpage

\begin{figure}
\psfig{file=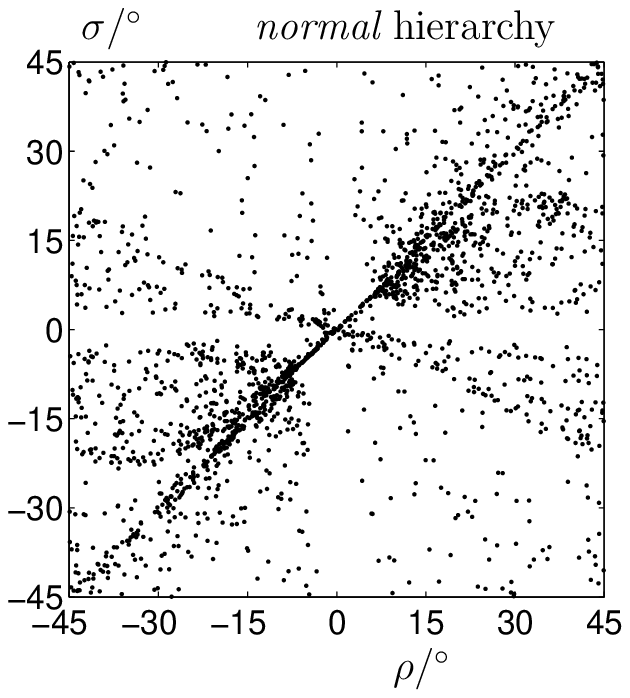, bbllx=2.2cm, bblly=6.0cm, bburx=12.2cm, bbury=16.0cm,%
width=12cm, height=12cm, angle=0, clip=0}
\end{figure}
\begin{figure}
\vspace{-3.5cm}
\psfig{file=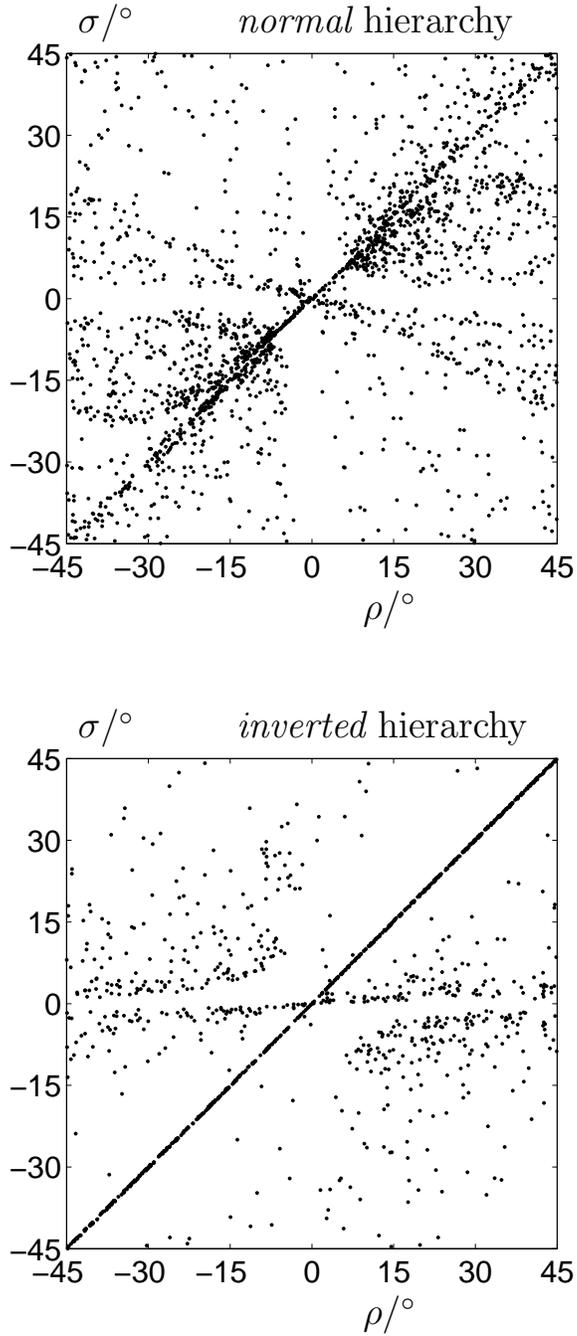, bbllx=2.2cm, bblly=6.0cm, bburx=12.2cm, bbury=16.0cm,%
width=12cm, height=12cm, angle=0, clip=0}\vspace{-2cm}\caption{The
parameter space of $\rho$ and $\sigma$ in \underline{\bf Scenario
(B)} with the {\it normal} or {\it inverted} neutrino mass
hierarchy. Note that the same parameter space can be obtained when
$\rho \rightarrow \rho \pm n\pi/2$ and $\sigma \rightarrow \sigma
\pm n\pi/2$ (for $n=1, 2, 3, \cdots$), as a straightforward
consequence of Eq. (17).}
\end{figure}


\begin{thebibliography}{99}
\bibitem{US} Note that one may also get $45^\circ \leq
\theta^{}_{23} \lesssim 52.2^\circ$, if $\theta$ takes negative
values from the third or fourth quadrant. Therefore,
$\theta^{}_{23}$ is in general allowed to vary in the range
$37.8^\circ \lesssim \theta^{}_{23} \lesssim 52.2^\circ$ for
$|\theta| \lesssim 24.6^\circ$. See the discussions below Eq. (28)
in section III.
\end{thebibliography}
\end{document}